\documentclass[%
 reprint, 
 nofootinbib, 
 amsmath,
 amssymb,
 aps, 
 ]{revtex4-1}
\usepackage{hyperref}
\usepackage{xcolor}

\newcommand{\mbf}{\mathbf}

\usepackage{dsfont}

\usepackage{diagbox}

\usepackage{supertabular}

\newcommand{\ra}{\rangle}
\newcommand{\la}{\langle}
\newcommand{\fij}{F_{ij}}
\newcommand{\hx}{x}
\newcommand{\hxi}{x_i}
\newcommand{\hxj}{x_j}
\newcommand{\hp}{p}
\newcommand{\hpi}{p_i}
\newcommand{\hpj}{p_j}
\newcommand{\hq}{q}
\newcommand{\hqi}{q_i}
\newcommand{\hqj}{q_j}

\newcommand{\hki}{k_i}
\newcommand{\hkj}{k_j}

\newcommand{\ijk}{\varepsilon_{ijk}}

\newcommand{\pdotq}{\mbf{\hp}\cdot\mbf{\hq}}

\newcommand{\commqk}{[\hqi,\hkj]=\ih\dij}
\newcommand{\commij}{[\hxi,\hpj]}
\newcommand{\commxx}{[\hxi,\hxj]}
\newcommand{\commpp}{[\hpi,\hpj]}
\newcommand{\ih}{i\hbar}
\newcommand{\dij}{\delta_{ij}}
\newcommand{\ie}{i.e., }
\newcommand{\eg}{e.g., }
\newcommand{\cf}{cf.\ }
\newcommand{\be}{\begin{equation}}
\newcommand{\ee}{\end{equation}}
\newcommand{\oo}[1]{\mathcal{O}({#1})}
\newcommand{\w}[1]{\widehat{#1}^{(d)}}
\newcommand{\ione}{{i_1 \dots i_d}}
\newcommand{\ionetwo}{{i_1 i_2 \dots i_d}}
\newcommand{\itwo}{{i_2 \dots i_d}}
\newcommand{\pone}{p_{i_1} \cdots p_{i_d}}
\newcommand{\ponetwo}{p_{i_1} p_{i_2} \cdots p_{i_d}}
\newcommand{\ptwo}{p_{i_2} \cdots p_{i_d}}
\newcommand{\C}{C}
\newcommand{\Cm}{C^{-1}}
\newcommand{\bs}[1]{\boldsymbol{#1}}
\newcommand{\bx}{\mbf{x}}
\newcommand{\bp}{\mbf{p}}
\newcommand{\bq}{\mbf{q}}
\newcommand{\bk}{\mbf{k}}
\newcommand{\bb}{\bs{\beta}}
\newcommand{\pbp}{\bp\cdot\bb\cdot\bp}
\newcommand{\pb}{\bp\cdot\bb}

\newcommand{\ba}{\bs{\alpha}}
\newcommand{\pa}{\ba\cdot\bp}
\newcommand{\bc}{\mbf{c}}

\newcommand{\pc}{\bp\cdot\bc}

\usepackage{soul}

\begin{document}

\preprint{APS/123-QED}

\title{On the algebraic approach to GUP in anisotropic space}

\author{Andr\'e Herkenhoff Gomes}
 \email{andre.gomes@ufop.edu.br}
\affiliation{%
 Departamento de F\'isica, Universidade Federal de Ouro Preto, Ouro Preto, MG, Brazil
}%


\begin{abstract}
Motivated by current searches for signals of Lorentz symmetry violation in nature and recent investigations on generalized uncertainty principle (GUP) models in anisotropic space, in this paper we identify GUP models satisfying two criteria: (i) invariance of commutators under canonical transformations, and (ii) physical independence of position and momentum on the ordering of auxiliary operators in their definitions. Compliance of these criteria is fundamental if one wishes to unambiguously describe GUP using an algebraic approach but, surprisingly, neither is trivially satisfied when GUP is assumed within anisotropic space. As a consequence, we use these criteria to place important restrictions on what or how GUP models may be approached algebraically.
\end{abstract}


\maketitle

\section{Introduction}
\label{sec:intro}

After facing the pitfalls of the old quantum theory for explaining atomic radiation, Heisenberg took the bold step of proposing the resolution to be of pure kinematic nature: finding the quantum quantity corresponding to the classical $x$ \cite{magic,magic-revealed}. Focusing on relating only observable quantities, this approach led him to elaborate the mathematical scheme for dealing with the quantum version of $x$, shortly after recognized by Born to satisfy $xp-px=\ih$ within the algebra of infinite-dimensional matrices \cite{born}. This canonical commutation relation then became the basic tenet from which Born, Heisenberg, and Jordan would develop a matrix theory for quantum phenomena \cite{born-jordan,born-heisenberg-jordan} --- later morphed into what is now regarded as the modern quantum theory after series of new insights, discoveries and contribution from many others \cite{longair}. As of today, clues to the long-standing search for a consistent quantum theory of gravity may be provided by investigation of \textit{modifications} of the canonical commutation relation as it provides an \textit{algebraic} approach to the generalized uncertainty principle (GUP) \cite{tawfik-diab-2014,tawfik-diab-2015}. Such modifications, generally written as 
\be\label{deformed}
[\hxi,\hpj]=\ih\dij \to \ih\fij(\bp),
\ee
introduce dependence of $F_{ij}\neq\dij$ on the particle's momentum and on parameters controlling departures from conventional quantum physics. These give rise to GUP through
\be
\Delta x_i \Delta p_j \ge \frac{1}{2}|\la \commij \ra | = \frac{\hbar}{2} |\la F_{ij} \ra|,
\ee
and encode (yet unknown) quantum effects of gravity on measurements of position and momentum. Specific modifications are usually motivated by results from more fundamental approaches to quantum gravitational phenomena, including: appearance of finite resolution for position measurements \cite{maggiore-algebra,kmm95,pedram12-plb2,mangano2016}, expected from model-independent arguments based on black hole thought experiments \cite{maggiore,micro-bh} and heuristically interpreted as a fundamental length scale in nature \cite{hossen2013}; and existence of classical regime at the Planck scale \cite{tsallis,petruzziello2020}, an idea dating back to ’t Hooft (\eg \cite{thooft}) and also predicted by independent models of crystal-like universe \cite{scardigli2010} or discrete time \cite{fadel-maggiore}.

Modifications of the canonical commutator under the hypothesis of spatial isotropy correspond to rotational covariant $\fij$, \ie a second-rank tensor build upon $\dij$ and $p_i p_j$ and functions of $\hp\equiv|\bp|$ only \cite{bruneton}. Among the models of GUP in isotropic space, the popular one proposed by Kempf sets $\fij = (1 + \beta\hp^2)\dij + \beta'\hpi\hpj$ and predicts a non vanishing minimum uncertainty of $\hbar\sqrt{3\beta+\beta'}$ for measurements of any spatial Cartesian coordinate \cite{kempf97}, where $\beta$ is a free parameter of the model and generally expected to be of order $(m_P c)^{-2}$ \cite{scardagli-lambiase-vagenas}, with $m_p$ the Planck mass. As mentioned before, this particular behavior is not a necessary hallmark of GUP, and for that reason a general framework for GUP in isotropic space was recently built and used to identify what the experimental bounds on particular GUP models mean for the general rotation invariant $\fij$ written as a power series on the momentum \cite{igup}. 

Thus, GUP introduces a fundamental length scale at the nonrelativistic regime. On the other hand, keeping invariant such scale becomes an important issue as soon as special relativity is taken into account. One approach to implement this is considering the $\kappa$-deformed Poincar\'e algebra \cite{kappa}. Even though it is an isotropic deformation of the standard algebra, such isotropy ought to be valid only at very specific inertial reference frames, being anisotropic at any boosted frame \cite{kost-mewes-photon}. Back at the nonrelativistic regime, this means that considering GUP in anisotropic space allows accessing a wider range of physical predictions other than those from very specific frames where isotropy holds. Very recently, first steps to consider GUP in this context were reported on \cite{nigup}. There, motivated by Lorentz symmetry violations as candidate signatures of quantum gravity \cite{tasson,bluhm}, one particular class of GUP models in anisotropic space is proposed and experimentally constrained within the framework of the Standard Model Extension \cite{sme1,sme2,sme3,datatable} --- possible experimental signals in the laboratory frame are placed around the 10 TeV scale \cite{nigup}.

The core idea to build GUP models in anisotropic space is to promote GUP parameters from constant scalars to constant tensors, \ie fixed background fields introducing anisotropies in space. Although the possibilities for such extension are many, we expect two basic criteria are satisfied should GUP be unambiguously described using an algebraic approach:
\begin{itemize}
    \item[(i)] Commutators $\commij$ and $\commxx$ are invariant under quantum canonical transformations --- although this is an automatic feature of conventional quantum mechanics, and found to be somewhat irrelevant in some cases even in the rotational covariant modified case, it is a nontrivial matter in the anisotropic scenario.
    \item[(ii)] Position and momentum may be expressed as functions of auxiliary operators $\hqi$ and $\hki$ satisfying $\commqk$, but we require the choice or ordering of these auxiliary operators in such functions to have no observable relevance --- interestingly, we show this is true whenever space is isotropic and generally not when it is anisotropic.
\end{itemize}
In this paper, our aim is to consider some representative anisotropic models and identify those satisfying the two criteria. It turns out, investigation of the second criterion fits well within the context of quantum canonical transformations too, hence it is the basic tool we employ all along this paper.

The paper is organized as follows. The anisotropic models considered in here are presented in Section \ref{sec:gup-nonisotropic}. Differently from the isotropic case, we notice right from the start that expressing position and momentum operators as functions of auxiliary operators requires extra care. These are dealt with in the context of quantum canonical transformations, discussed in Sec.\ \ref{sec:intro-canonical-transf}. A brief interlude is made in Sec.\ \ref{sec:symmetricity} to derive explicit expressions for the position operator and its canonical transformation that are used in Sec.\ \ref{sec:commxx}, where invariance of commutators is investigated, and in Sec.\ \ref{sec:operator-ordering}, where canonical transformations are used to determine whether expressing position and momentum operators as functions of auxiliary operators is viable for the models we investigate. Finally, in Sec.\ \ref{sec:overlap} we summarize the models meeting the two criteria discussed above. We close this work with our concluding remarks in Sec.\ \ref{sec:conclusion}.

\section{Anisotropic Models}
\label{sec:gup-nonisotropic}

In the algebraic approach to GUP, setting the modification (\ref{deformed}) to the canonical commutation relation,
\be
\label{deformed-commxp}
\commij = \ih \fij(\bp),
\ee
immediately introduces a superior algebraic complexity when compared to the standard commutator simply proportional to $\dij$. One profitable approach to circumvent this extra challenge is to express position $x_i$ and momentum $p_i$ operators as functions of auxiliary operators $q_i$ and $k_i$ that satisfy the conventional algebra
\be
\commqk \qquad \text{and} \qquad [q_i,q_j]=[k_i,k_j]=0.
\ee
Determination of $\hxi(\bq,\bk)$ and $\hpi(\bq,\bk)$ such that (\ref{deformed-commxp}) is fulfilled also depends on what the commutators $\commpp$ and $\commxx$ are. In particular, we set the first to zero,
\be\label{commpp}
\commpp=0,
\ee
to keep the translation group as conventional, which imposes $\hpi=\hpi(\bk)$ once we require $\hpi\to\hki$ in the limit where conventional quantum mechanics is recovered. This allows fulfilling (\ref{deformed-commxp}) by writing $\hxi$ as the sum of $\fij(\bp)\hqj$ and some function $F_i(\bp)$ that does not affect the commutator of position and momentum (this function is discussed in Sec.\ \ref{sec:symmetricity}). At last, with $\hxi(\bq,\bk)$ at hand, $\commxx$ can be computed --- notice dependence of $x_i$ on the $k_i$ suggests $\commxx\neq0$ for general $F_{ij}$, which is confirmed by use of the Jacobi identity $[x_i,[x_j,p_k]]+\text{cyclic}=0$ indeed.

Although position and momentum may be formally represented by any suitable $\hxi(\bq,\bk)$ and $\hpi(\bk)$ satisfying the above commutators, an often simplifying approach is to exploit the \textit{freedom} we have in realizing this algebra in the Hilbert space of momentum wave functions, where $\hpi$ is represented by the usual multiplicative operator and the choice
\be\label{pk}
\hpi = \hki
\ee
can be made. As a consequence, $\hxi(\bq,\bp)$ is represented by a modified derivative operator based on
\be
\hqi=\ih\partial_{\hpi} \equiv \ih\frac{\partial}{\partial p_i}.
\ee
Since representation on position wave functions is known to be unachievable whenever (\ref{deformed-commxp}) implies existence of non vanishing minimum spatial resolution \cite{kmm95,pasquale-xbasis}, representation on momentum wave functions is preferred in order to have $\fij$ as general as possible.

In this scheme, for isotropic space, $\fij$ in its more general form is written as
\begin{equation}
\label{isotropic}
\fij(\bp) = f(\hp)\dij + g(\hp)\hpi\hpj,
\end{equation}
where $f$ and $g$ are functions of $\hp\equiv|\bp|$ only. Giving up isotropy allows $\fij$ to have virtually any tensor structure depending on what sort of anisotropies are considered. To be concrete, in this work we consider
\begin{equation}
\label{Fij}
    \fij(\bp) = f(\bp)\dij + g_i(\bp)\hpj + \hpi h_j(\bp)
\end{equation}
with
\begin{align}\label{particular-anisotropies}
    f & = 1 + \boldsymbol{\alpha}\cdot\bp + \bp\cdot\boldsymbol{\beta}\cdot\bp + \cdots,
    \nonumber\\
    g_i & = \alpha'_i + (\bp\cdot\boldsymbol{\beta'})_i + \cdots,
    \nonumber\\
    h_j & = c_j + (\bp\cdot\mbf{d})_j + \cdots,
\end{align}
where $\ba$, $\ba'$, and $\bc$ are constant vectors while $\boldsymbol{\beta}$, $\boldsymbol{\beta}'$, and $\mbf{d}$ are constant second-rank tensors, all having real components and playing the role of fixed background fields breaking spatial isotropy; in particular, vector anisotropies and any of odd tensor rank are associated to parity- and time reversal-violating behavior as well. Ellipsis comprises couplings of other background tensors to higher powers on the momentum and will not be considered explicitly. Notice we do not claim to have exhausted all possibilities for the structure of $\fij$. The above choice (\ref{Fij}) aims a simple rotation \textit{non} covariant extension of the rotation covariant $\fij$ in (\ref{isotropic}) and one suitable for inspecting novelties coming from, \eg $\fij \neq F_{ji}$. Additionally, whenever we consider at most \textit{one} anisotropy in $f$, $g_i$, and $h_i$, we may write these three as
\begin{align}\label{single-anisotropies}
    f & = 1 + \w{f} \equiv 1 + f_\ionetwo \ponetwo,
    \nonumber\\
    g_i & = \w{g}_i \equiv g_{i \itwo} \ptwo,
    \nonumber\\
    h_j & = \w{h}_i \equiv h_{i \itwo} \ptwo,
\end{align}
where the number $d$ on each of them may be set independently. The wide hat notation abbreviates the tensorial indices of anisotropies, where the \textit{total} number of indices is revealed by the superscript  ($d>0$), and also abbreviates the contraction with momentum factors --- for instance, the symbol $\w{g}$ represents the contraction $\w{g}_i p_i = g_\ione \pone$. To the rest of this paper, this notation is always linked to considering at most one kind of anisotropy in each $f$, $g_i$, and $h_j$.

Having extended the algebraic approach to GUP to anisotropic scenarios, we are faced with some interesting conceptual questions right from the start. In particular, $\fij$ is generally no longer the same as $F_{ji}$, and since the position operator depends on the contraction $\fij\hqj$ (Sec.\ \ref{sec:symmetricity}), we may ask what fundamentally different properties models with $g_i$ or $h_i$ above may exhibit. Another, more subtle question relates to the choice of expressing $\hxi$ in terms of $\fij(\bp)\hqj$ or $\hqj \fij(\bp)$, or even something in between. Should we care about the ordering of auxiliary operators? This question also exists within the algebraic approach to GUP in isotropic space, although rather unimportant after all as we show in Sec.\ \ref{sec:operator-ordering}, but it is highly nontrivial when allowing for space anisotropies.

In what follows, quantum canonical transformations are used to investigate these and other related questions. Right from the onset the choice (\ref{pk}) of $\hpi=\hki$ as the conventional multiplicative operator is made to satisfy $\commpp\equiv0$. Although this choice of representation limits the available set of canonical transformations to those keeping $\hpi$ invariant, it amounts to no physical loss of generality and suffices for our goals.

\section{Quantum canonical transformations}
\label{sec:intro-canonical-transf}

Quantum canonical transformations $\C(\bx,\bp)$ are identified as commutator preserving transformations \cite{born-heisenberg-jordan}, which basically demands existence of suitable $\Cm(\bx,\bp)$. Requiring that the transformation connects only physically equivalent systems further constrains it to be isometric, \ie a norm-preserving isomorphism between Hilbert spaces; nor $\C$ or $\Cm$ shall annihilate physical states nor affect state normalization \cite{anderson1,anderson2}. Here we briefly review and discuss what such transformations mean for commutators and expectation values in the context of this paper.

\subsection{Commutators}
\label{sec:commutators}

Quite generally, the action of a canonical transformation $\C(\bx,\bp)$ on an operator $M(\bx,\bp)$ and a state vector $|\psi\rangle$ is
\be
    M' = \C M \Cm
    \quad \text{and} \quad 
    |\psi'\rangle = \C |\psi\rangle.
\ee
The set of canonical transformations keeping $\hpi=\hki$ as the usual multiplicative operator is restricted to those depending only on the momentum, \ie $C=C(\bp)$. From now on, we deal exclusively with such set.

For conventional quantum mechanics, the \textit{covariant} behavior of $[\hxi,\hpj]=\ih\dij$ under such canonical transformations,
\be
[\hxi',\hpj'] = \C[\hxi,\hpj]\Cm = \ih\dij = [\hxi,\hpj],
\ee
further reduces to an \textit{invariance} for any acceptable $C(\bx,\bp)$. Quantum mechanics with modified canonical commutation relation behaves quite differently because
\be
[\hxi',\hpj'] = \C\commij\Cm = \ih\C \fij(\bp)\Cm
\ee
is merely covariant for general $C(\bx,\bp)$. Is this an issue? We skip this question for now because sticking with $\hpi=\hki$ as the conventional multiplicative operator ends up enforcing invariance since, in this case, $C=\C(\bp)$.  Furthermore, since the commutator of position operators need no longer vanish, we note the transformation
\be [\hxi',\hxj'] = C \commxx C^{-1} = \commxx + C [ \commxx,C^{-1} ]
\ee
has to be investigated as well, but we defer it to Sec.\ \ref{sec:commxx}. Notice that by simply allowing for $\commij\neq\ih\dij$ the basic commutators already transform non trivially under canonical transformations.

\subsection{Expectation values}

Expectation values, in contrast, are generally \textit{not} invariant even for conventional quantum mechanics since they may transform non unitarily,
\be
\langle\psi|M|\phi\rangle = \langle\psi'|(\C\C^\dagger)^{-1}M'|\phi'\rangle \neq \langle\psi'|M'|\phi'\rangle.
\ee
This is not a problem, of course, as it merely changes the integral measure $d\mu$ in the definition of the scalar product to $d\mu/(CC^\dagger)$. The expression is not invariant, but the scalar product \textit{per se} is preserved, \ie the transformation is an isometry.

As a simple illustration, consider the effect of a canonical transformation based on some $C(\bp)$. The momentum operator is unaffected and, for simplicity, its spectral basis will be used for representation in momentum-space.  The position operator $\hxi$ transforms into
\begin{align}\label{canonical-transf-x}
\hxi'
& = \hxi + C[\hxi,C^{-1}] \nonumber\\
& = \hxi + \fij C[\hqj,C^{-1}] \nonumber\\
& = \hxi + \ih \fij C\partial_{j}C^{-1} \nonumber\\
& = \hxi - \ih \fij \partial_{j}\ln{C},
\end{align}
where the simpler notation $\partial_i\equiv \partial_{p_i}$ will be adopted along this paper. The contribution from the transformation is momentum-dependent only and expresses the common wisdom that any such function can be added to the position operator without changing the canonical commutation relation between position and momentum --- whether the commutator of position operators changes in our context is to be investigated (Sec.\ \ref{sec:commxx}). There is more information here, though, as we discuss in the rest of this section.

The term $\ih \fij\partial_i \ln{C}$ above is purely real if $\C(\bp)$ is an unitary operator, say $\C=e^{i\Gamma}$ with real $\Gamma(\bp)$. Such $\C$ can be traced back to an unobservable change of basis vectors. For instance, suppose two sets of momentum eigenstates are related by  $|\mathfrak{p}\rangle=C(\bp)|\bp\ra$. In momentum-space representation, the expectation value of the position operator reads
\begin{align}
\langle\psi|\hxi|\phi\rangle 
& = \int d^3p\, \langle\psi|\bp\rangle\langle\bp|\hxi|\phi\rangle \nonumber\\
& = \int d^3p\, \langle\psi|C^\dagger|\mathfrak{p}\rangle\langle\mathfrak{p}|C\hxi|\phi\rangle \nonumber\\
& = \int d^3p\, \langle\psi'|\mathfrak{p}\rangle\langle\mathfrak{p}|C\hxi C^\dagger|\phi'\rangle \nonumber\\
& = \int d^3p\, \langle\psi'|\mathfrak{p}\rangle\langle\mathfrak{p}|\hxi'|\phi'\rangle \nonumber\\
& = \langle\psi'|\hxi'|\phi'\rangle.
\end{align}
The unitary change on the basis vector induces a change on both state vectors and the position operator, or any other operator for what matters, with an overall cancellation leaving the scalar product invariant. 

Conversely, $\ih \fij\partial_i \ln{C}$ is purely imaginary if $\C(\bp)$ is a non-unitary operator, \eg $\C=e^\Gamma$ with real $\Gamma(\bp)$. The freedom to add an imaginary function of momenta to $\hxi$ as in (\ref{canonical-transf-x}) comes with the price of changing the integral measure in the definition of the scalar product:
\begin{align}
\label{definition-exp-value-c}
\langle\psi|\hxi|\phi\rangle 
& = \int d^3p\, \psi^*(\bp)\hxi\phi(\bp) \nonumber\\
& = \int d^3p\, [C^{-1} \psi'(\bp)]^* [C^{-1} \hxi' C] [C^{-1}\phi'(\bp)] \nonumber\\
& = \int \frac{d^3p}{C^2}\, \psi'^*(\bp)\hxi'\phi'(\bp) \nonumber\\
& = \langle\psi'|\hxi'|\phi'\rangle_C,
\end{align}
where the subindex is a reminder the definition of the scalar product for the transformed quantities is to be taken under the integral measure $d^3p/C^2$. This conclusion is particularly relevant for establishing a relation between canonical transformations and the symmetricity of $\hxi$ (Sec.\ \ref{sec:symmetricity}).

\subsection{Covariance or invariance?}

For conventional quantum mechanics, the identification of canonical transformations as those preserving canonical commutators is unambiguous as covariance actually reduces to invariance; the relation among certain operators follows the same mathematical structure and physical intuition. At the same time, we can argue there is no fundamental issue with non canonical transformations; after all, modeling an experimentally accessible quantum system with known observables involves constructing an adequate Hamiltonian, and transforming it in different ways, be it canonically or not, may be useful for gaining physical insights, simplifying calculations, and so on.\footnote{To illustrate this point, consider the one-dimensional quantum harmonic oscillator. The transformation from canonically conjugated position and momentum $(x,p)$ to ``ladder'' operators $(a,a^\dagger)$ is non canonical because $[a,a^\dagger]=1\neq \ih = [x,p]$. Nevertheless, its helps providing an algebraic solution for this system.}

For models with modified commutators, the question arises: should we be satisfied with covariance or should we further restrict the set of canonical transformations to those also leading to invariance? As long as the transformation $C(\bx,\bp)$ is isometric, there seems to be no physical reason motivating such restriction. On the other hand, restricting the set of $C$ to those leaving invariant the commutator algebra of position and momentum has the advantage of helping sorting out set of models with exact same behavior --- \eg identifying whether different proposals for representing the position operator, $\hxi \sim \fij\hqj$ or $\hxi \sim \hqj \fij$, belong to models with identical mathematical structure.

Here we adopt a pragmatic approach by all means. For simplicity, we opted to maintain the momentum operator conventional at all circumstances, \ie $\hpi'=\C\hpi\Cm=\hpi$, restricting our considerations to canonical transformations with $C=C(\bp)$ only. Incidentally, such transformation also leave $\commij$ invariant, as seem before. To what extent this is also true for $\commxx$ still has to be answered (Sec.\ \ref{sec:commxx}), but for that we need first an explicit expression for $\hxi$ in terms of $\hqi$ and $\hpi$. This is done in the next section after requiring $\hxi$ to be a symmetric operator.

\section{Symmetricity and canonical transformations}
\label{sec:symmetricity}

An essential requirement for models with a generalized uncertainty principle derived from a modified algebra for $\hxi$ and $\hpi$ is the symmetricity of the position operator, \ie $\langle\psi|\hxi|\phi\rangle = \langle\phi|\hxi|\psi\rangle^\ast$. This ensures position eigenvalues are real even in the case $\hxi$ is not self-adjoint, \eg due to the existence of a non vanishing minimum uncertainty on its measurement \cite{kmm95,kempf2000,pedram12-prd}. The symmetricity of the momentum operator, and actually its self-adjointness, is an immediate result once it can be represented as the usual multiplicative operator. The same cannot be said of $\hxi$, thus we devote this section to ensure its symmetricity and establish a link to canonical transformations.

We start with a simple investigation. A seemly fair modification on the position operator sufficient to implement (\ref{deformed}) is $\hxi \to \fij\hqj$ as it gives $[F_{ik}\hq_k,\hpj]=\ih \fij$. Investigating its symmetricity under conventional definition for the scalar product in momentum space representation, we get
\be\label{trivial-case}
\langle\psi|\fij\hqj|\phi\rangle = \langle\phi|\fij\hqj|\psi\rangle^* -\ih\int d^3p (\partial_j \fij) \psi^*\phi.
\ee
Such modification of $\hxi$ is symmetric for all $\psi$ and $\phi$ only for the trivial case $\fij=\dij$, where the last term above vanishes, but this would be just conventional quantum mechanics.

To go beyond and really implement a generalization of the canonical commutator, the term $\fij\hqj$ above must be supplemented by a function $F_i(\bp)$ of the momentum alone, which reminds us of the canonical transformation (\ref{canonical-transf-x}) associated with $C(\bp)$. Thus, we consider $ \hxi' = \fij\hqj + \ih F_i$ and deal with the canonically transformed position operator $\hxi'$ for greater generality. This extra momentum-dependent term can be suitably chosen to make $\hxi'$ a symmetric operator; notice
\begin{align}
\langle\psi'|\hxi'|\phi'\rangle_C = & \langle\phi'|\hxi'|\psi'\rangle^*_C \nonumber\\
& -\ih\int d^3p \left[ \partial_j \left( \frac{\fij}{C^2}\right) - \frac{2F_i}{C^2} \right]\psi'^*\phi'
\end{align}
suggests $\hxi'$ is symmetric if we set $F_i = \frac{1}{2} C^2\partial_j ( \fij C^{-2} ) = \frac{1}{2}\partial_j \fij - \fij \partial_j \ln C$ indeed.

At last, comparing the general form of the symmetric position operator we have just found,
\be\label{hxi-prime}
\hxi' = \fij\hqj + \tfrac{1}{2}\ih\partial_j \fij - \ih \fij\partial_j\ln{C},
\ee
to the canonical transformation (\ref{canonical-transf-x}) of $\hxi$ under $\C(\bp)$, we identify
\be\label{hxi}
\hxi = \fij\hqj + \tfrac{1}{2}\ih\partial_j \fij,
\ee
which of course is $\hxi'$ when no transformation is implemented ($C=1$). These expressions reveal the structure of the position operator in quantum theories with a modified canonical commutation relation of the sorts of (\ref{deformed-commxp}) and its transformation under canonical transformations with $\C(\bp)$. As mentioned before, these are the ones leaving the commutator (\ref{deformed-commxp}) invariant and, in the next section, we investigate under what circumstances the same can be said of the commutator of position components.

To close this section, we briefly recover from the above some general expressions for GUP in isotropic space, where $\fij=f\dij+g p_i p_j$. Setting $C^2=(f+gp^2)^{1-\varepsilon}$, we can recast (\ref{hxi-prime}) as
\be
\hxi'= \ih( f\dij + g p_i p_j)\partial_{p_i} + \ih\gamma\hpi,
\ee
with $\gamma=\gamma(p)$ defined by the relation
\be
\varepsilon = \frac{2p(\gamma-g)}{\partial_p f+(\partial_p g)p^2+2gp},
\ee
where $\partial_p f \equiv d f/d p$ and similarly for $g$ since these are functions of $p=|\bp|$ only. Scalar products are then expressed as
\be\label{scalarproduct}
\langle\psi'|\phi'\rangle_\varepsilon = \int \frac{d^3p}{(f+gp^2)^{1-\varepsilon}}\psi'^*(\bp)\phi'(\bp),
\ee
where $\varepsilon=1$ means no canonical transformation is performed and the integral measure is just $d^3p$. These are well-known expressions in the literature, but serve here as a consistency check of our approach using canonical transformations.

One might hope for a simple generalization of (\ref{scalarproduct}) for the case of GUP in anisotropic space, but we could find none at least for $F_{ij}$ given by (\ref{Fij}), seemly due to the complicated structure that comes from the anisotropic generalization of $g\hpi\hpj$. The sole exception is the particular case $\fij=f(\bp)\dij$ with $f$ given by (\ref{particular-anisotropies}), where setting $C^2=f^{1-\varepsilon}$ allows for a similar expression for $\hxi'$ above but with $\gamma\hpi$ replaced by $\tfrac{1}{2}\varepsilon\partial_i f$ and the same expression for the scalar product but with $g=0$ instead.

\section{Commutator of position operators}
\label{sec:commxx}

At this point, we have seen the commutator of position and momentum (\ref{deformed-commxp}) is generally invariant only under canonical transformations with $C(\bp)$ if with choose the momentum operator as the usual multiplicative operator, and that such transformations also keep intact the symmetricity of $\hxi$. Finally, here we investigate the effects of such transformations on the commutator of position components.

We begin using (\ref{hxi-prime}) to compute the commutator directly:
\begin{align}
\label{deformed-commxx}
& \commxx
\nonumber\\
& = F_{im}[\hq_m,F_{jn}]\hq_n + \tfrac{1}{2}\ih[F_{im}\hq_m,\partial_n F_{jn}] - (i \leftrightarrow j)
\nonumber\\
& = \ih F_{im}\left[ (\partial_m F_{jn})\hq_n + \tfrac{1}{2}\ih \partial_m\partial_n F_{jn} \right] - (i \leftrightarrow j)
\nonumber\\
& = \ih \varepsilon_{tij}\varepsilon_{tkl} F_{km}\left[ (\partial_m F_{ln})\hq_n + \tfrac{1}{2}\ih \partial_m\partial_n F_{ln} \right],
\end{align}
where we used $\varepsilon_{tij}\varepsilon_{tkl}=\delta_{ik}\delta_{jl}-\delta_{il}\delta_{jk}$ for the last equality. Allowing for a canonical transformation with $C(\bp)$, it transforms covariantly to
\begin{align}
\label{transf-commxx}
[\hxi',\hxj'] 
& = \commxx + C [ \commxx,C^{-1} ] \nonumber\\
& = \commxx + \ih \varepsilon_{tij}\varepsilon_{tkl} F_{km} (\partial_m F_{ln}) C [\hq_n,C^{-1} ] \nonumber\\
& = \commxx +\hbar^2 \varepsilon_{tij}\varepsilon_{tkl} F_{km} (\partial_m F_{ln}) \partial_n \ln C.
\end{align}
Invariance, therefore, requires the vanishing of the second term on the right-hand side of the last equality above. In what follows, we investigate this ``extra'' term first for isotropic models and then for anisotropic ones.

\subsection{Isotropic models}

For the rotation covariant $\fij$ (\ref{isotropic}), the commutator (\ref{transf-commxx}) is actually \textit{invariant} as long as $C$ has no direction dependence, meaning $C=C(p)$. To see this explicitly, first notice the derivative $\partial_i$ of some scalar function $\phi(p)$ can be rewritten as $\partial_i\phi = \phi' p_i p^{-1}$ with primes here denoting derivative with respect to $p$. The extra term in (\ref{transf-commxx}),
\begin{align}
& F_{km}(\partial_m F_{ln}) \partial_n \ln C
\nonumber\\
& = f[(f' + g' p^2)p_k p_lp^{-1} + g( \delta_{lk} p^2 + p_k p_l)]p^{-1}C^{-1}C'
\nonumber\\
& \quad + g (f'+ g'p^2 + 2gp] p_k p_l C^{-1}C',
\end{align}
is then found to be symmetric under $k\leftrightarrow l$; hence, it vanishes identically when contracted with $\varepsilon_{tkl}$ as in (\ref{transf-commxx}). This conclusion was to be expected on the grounds of spatial isotropy: any rotation covariant tensor $I_{kl}(\bp)$ can be written as a combination involving only $\delta_{kl}$ and $p_k p_l$, resulting in $\varepsilon_{tkl}I_{kl}\equiv 0$.

\subsection{Anisotropic models}

For anisotropic space and $\fij$ given by (\ref{Fij}), our analysis suggests invariance of $\commxx$ is generally \textit{not} attainable at least for any simple choice of $C$. The conclusion whether the extra piece in (\ref{transf-commxx}) vanishes, or what conditions are needed for that, depends on the anisotropies considered and if the calculations are handled exactly or to some specific power on the anisotropies. Since any possible anisotropy is expected to be very small compared to any attainable scale on current or near-future experiments, analysis based on perturbative approaches may seem to be the obvious choice, but this need not always be the case; for instance, whenever one envisages possible applications to designed condensed-matter systems with large anisotropies that might serve as analogue models of GUP-based quantum mechanics.

\subsubsection{Non-perturbative approach}
\label{sec:symm-x-non-pert}

A first case is that of $\fij = f(\bp)\dij$, with $f$ as general as possible and not only of the particular form in (\ref{particular-anisotropies}). The extra term in the commutator (\ref{transf-commxx}) then simplifies to $\varepsilon_{tmn}f \partial_m f \partial_n\ln C$ and vanishes for any of the two distinct choices: $\ln C$ or $C$ proportional to some function of $f$ with the further restriction $\C\to1$ for $p\to0$ to avoid issues with the existence of $\Cm$ or state normalization. As far as leaving the commutator (\ref{transf-commxx}) invariant, both choices are equally acceptable.

For models of $f=1$ and $g_i(\bp)$ with only $\alpha'_i$ or $\beta'_{ij}$ in (\ref{particular-anisotropies}), invariance of the commutator is found for $C$ proportional to some function of $\ba'\cdot\bp$ or $\bp\cdot\bb'\cdot\bp$ with additional restriction $\beta'_{ij}=\beta'_{ji}$, respectively. More generally, we can set $g_i=\w{g}_i$ as in (\ref{single-anisotropies}) with anisotropy $g_\ione$ symmetric under permutation of any pair of indices and come to the conclusion there is $C$ proportional to some function of $\w{g}$ leaving $\commxx$ invariant under the canonical transformation.

Aside from the above case with general $f$, we identified only one model with multiple independent anisotropies allowing for invariance of the position commutator; namely, one with vector anisotropies $\ba$ and $\ba'$ for $C\propto \exp[-\frac{1}{2} + \frac{1}{2}(1+\pa)^2-\ba'\cdot\bp]$. If we otherwise assume proportionality among any of the set $\alpha_i \alpha_j$, $\alpha'_i \alpha'_j$, $\beta_{ij}$, and $\beta'_{ij}$, we find there is always $C$ rendering the commutator invariant as long as $C$ is proportional to any reasonable function of the contraction of momentum and the only anisotropy being considered as independent. This conclusion can be extended to any model with $f$ and $g_i$ given by (\ref{single-anisotropies}) as long as the relation $g_\ione \propto f_\ione$ is assumed.

In contrast, there is no $C$ leaving $\commxx$ invariant for any model based on $\fij=\dij + p_i h_j$. This conclusion comes after noticing that the extra term in $[\hxi',\hxj']$,
\begin{align}
\varepsilon_{tkl} p_k ( h_l h_n - \partial_l h_n ) \partial_n \ln\C,
\end{align}
does not vanish because $\mbf{h} h_n- \bs{\partial}h_n$ is neither vanishing nor orthogonal to $\bp$ for arbitrary momentum directions. Although unexpected at first, insight on this result may be provided noticing anisotropies are physical features of space but, for models with $h_i \neq 0$, they couple only to the auxiliary, possibly devoid of observable significance, operator $\bq$ in the expression (\ref{hxi}) for position operator, \ie $\hxi \supset \hpi\mbf{h}\cdot\bq$. This might suggest this particular anisotropic model is physically unmotivated or that the algebraic approach with auxiliary operators as defined in Sec.\ \ref{sec:gup-nonisotropic} is unsuitable in this case; discussion at the end of Sec.\ \ref{sec:ordering-aniso} reinforces this point of view.

\subsubsection{Perturbative approach}
\label{sec:symm-x-pert}

Our statements so far are valid considering computations that are non-perturbative on the anisotropies. Surely any of the discussed models has to break down for higher momenta as it may approach the so far unknown quantum gravity regime. In this sense, due to the expected very small value of possible anisotropies, such models may be well approximated to first order on them. To begin with, consider a model with only a single anisotropy symbolically denoted by $a$ and corresponding to one term out of those in (\ref{particular-anisotropies}) or (\ref{single-anisotropies}); then we can write $\fij=\dij+f_{ij}(\bp)$ where the anisotropic part $f_{ij}$ is $\oo{a}$. Since $\partial_m F_{ln} \sim \oo{a}$, the extra contribution from canonical transformation to $\commxx$ in (\ref{transf-commxx}) is only $\oo{a^2}$ for any $C$ satisfying $\partial_n\ln C \sim \oo{a}$, which can be always be achieved. Extension of this argument for the case of multiple anisotropies is immediate. To this approximation, canonical transformations leaving invariant all commutators discussed so far can be immediately found, including the case $h_i(\bp)\neq0$ at the end of the previous section.

\section{Operator ordering}
\label{sec:operator-ordering}

For mathematical easiness, $x_i$ has been expressed in terms of the auxiliary $\hqi$ and $\hki=\hpi$ defined in Sec.\ \ref{sec:gup-nonisotropic}, see (\ref{hxi}). Operator $\hki$ acquired the physical meaning of momentum after we exploited the freedom to make such identification while still respecting the commutator algebra of $\hxi$ and $\hpi$. Despite its usefulness, speaking in general, both $q_i$ and $k_i$ should be physically dispensable tools in the sense that the physics of GUP is defined by the algebra of position and momentum only. In this context, the particular ordering of $\hqi$ and $\hki$ in $\hxi(\bq,\bk)$ is expected to be \textit{unobservable}. Therefore, writing down these auxiliary operators in different orderings must be equivalent to performing canonical transformations on $x_i$.

To investigate whether any reordering of $q_i$ and $k_i$ (equivalently, $p_i$) is really unobservable, we need to care only for the term $\fij\hqj$ inside $\hxi(\bq,\bp)$ as given by (\ref{hxi}). In particular, this ordering of $F_{ij}$ and $q_j$ is a \textit{definition} from which $\hxi(\bq,\bp)$ was constructed. If defined differently, as $\hqj \fij$ or with $\hqj$ among momentum factors in $\fij$, we could commute it back to $\fij\hqj$ at the price of having extra terms depending on the momentum alone. Due to these extra terms, the result would not be $\hxi$ as originally defined, but if we expect such redefinition to be unobservable, it should be equivalent to $\hxi'$ derived from the canonical transformation (\ref{hxi-prime}) with an appropriate $C(\bp)$.

Next, we consider the effects of the mentioned reordering first for GUP in isotropic space and then in anisotropic space.

\subsection{Isotropic models}

For a simpler picture first, consider just moving $\hqi$ to the left of $\fij(\bp)=f(p)\dij+g(p) p_i p_j$ in the expression (\ref{hxi-prime}) for $\hxi'$. This rearrangement introduces
\be
[\fij,\hqj] = -\ih(f'+g'p^2+4gp) \frac{p_i}{p}.
\ee
Is such change observable or can it be made unobservable by a compensating canonical transformation? Comparing (\ref{hxi-prime}) with (\ref{hxi}) we notice a canonical transformation with $C(p)$ adds to $\hxi$ a contribution of
\be
-\ih \fij\partial_j\ln C = - \ih (f+gp^2)  \frac{d(\ln C)}{dp} \frac{p_i}{p}.
\ee
Thus, choosing
\be\label{canonical-reordering}
C(p) = \exp \left\{
-\int dp \frac{f'+g'p^2+4gp}{f+gp^2}
\right\}
\ee
the rearrangement we described is made unobservable. It is straightforward to extend this conclusion to any intermediate reordering. For instance, consider the particular case $\hxi \supset p^{m+n}\hqi$.  Passing $n$ momentum factors to the right of $\hqi$ produces a polynomial in the momentum like
\begin{align}
p^m \left[ p^n, \hqi \right]
= -\ih n p^{m+n-1} \frac{p_i}{p}.
\end{align}
This extra polynomial is correspondingly compensated by a canonical transformation like (\ref{canonical-reordering}), but replacing $f'+g'p^2+4gp$ there by $n p^{m+n-1}$.

For the isotropic scenario, we conclude that, as long as the integral in (\ref{canonical-reordering}) exists and leads to acceptable $C(p)$, there is no observable consequence coming from ambiguities on the ordering of auxiliary operators in the definition of the position operator.\footnote{Analogous conclusion is found on a different context on \cite{bosso-luciano} for the one-dimensional commutator $[x,p]=\ih f(p)$.}

\subsection{Anisotropic models}
\label{sec:ordering-aniso}

The simple result we find for the isotropic case cannot be generally extended to the anisotropic case based on $\fij$ as given by (\ref{Fij}). Ordering independence seems to be model-dependent and should be investigated on a case-by-case basis. For any particular $\fij$, the general idea is to start passing $\hqi$ to the left of $\fij(\bp)$ in the expression (\ref{hxi-prime}) for $\hxi'$, generating a polynomial in the momentum,
\be\label{added-from-reodering}
[\fij,\hqj] = - \ih \partial_j \fij.
\ee
Next, noticing from (\ref{hxi-prime}) that, if there is $C(\bp)$ satisfying
\be\label{condition-reordering}
\partial_j \fij + \fij \partial_j \ln C = 0,
\ee
then the extra polynomial in the momentum can be compensated by the canonical transformation and the reordering $\fij\hqj \to \hqj \fij$ in $\hxi$ is unobservable --- otherwise, different orderings represent inequivalent models. At last, one might want to verify if the same can be done for intermediate reordering, \ie placing $\hqi$ somewhere in between the momentum factors of $\fij(\bp)$. If so, ordering of auxiliary operators in the definition of the position operator is completely immaterial.

As in the previous section, we report our conclusions for both non-perturbative and perturbative approaches.

\subsubsection{Non-perturbative approach}

Starting with the case $\fij=f(\bp)\dij$, we find condition (\ref{condition-reordering}) is satisfied for $C(\bp) \propto f^{-1}(\bp)$. That means the choices $\fij\hqj$ and $\hqj \fij$ for $\hxi$ are physically equivalent in this model. In contrast, intermediate reordering do not seem to be compensated by any canonical transformations unless $f$ is isotropic (\cf previous section) or has a single anisotropic term, \ie $f=1+\w{f}$ as in (\ref{single-anisotropies}); only in this case we found $C\propto f^{-n}$ renders intermediate reordering unobservable for some suitable choice of $n$. In particular, this is the case of models with only $\alpha_i$ or $\beta_{ij}$.

Moving on to models based on $\fij = \dij + g_i(\bp)\hpj$, whenever only a single anisotropic term is considered, hence $g_i=\w{g}_i$, we find any reordering of auxiliary operators is made unobservable for $C(\bp)=(1+\w{g})^n$ with $g_\ione$ symmetric under exchange of any pair of indices. This includes models with only $\alpha'_i$ or $\beta'_{ij}=\beta'_{ji}$.

We also find reordering is unobservable for models where $f$ and $g_i$ of (\ref{single-anisotropies}) have anisotropies related by $g_\ione = \kappa f_\ione$, for which $C(\bp) = [ 1 + (1+\kappa) \w{f} ]^{ -\frac{n}{1+\kappa} }$ compensates any reordering. Particular cases include models with $\alpha'_i=\kappa\alpha_i$ or $\beta'_{ij}=\kappa\beta_{ij}$ only.

The situation is quite the opposite for models with $\fij = \dij + p_i h_j$ as there is no $C(\bp)$ satisfying (\ref{condition-reordering}). Definition of $\hxi$ in terms of $\fij\hqj$ or $\hqj \fij$ corresponds, at this level of investigation at least, to inequivalent models. To illustrate the reason, consider a model with only $c_{i}$ in (\ref{particular-anisotropies}). Up to a factor of $-\ih$, reordering adds the contribution $[\fij,\hqj] \sim c_i$ to $\hxi$ while a canonical transformation contributes $\fij\partial_j\ln C = \partial_i\ln C + p_i \bc\cdot\mbf{\partial} \ln C$. These cannot be adjusted to compensate each other due to their generally differing directions (one only along $c_i$ and the other also along $\hpi$). This conclusion can be extended to arbitrary $h_j$: reordering would be unobservable if there is $C$ such that (\ref{condition-reordering}) is respected, \ie for $\partial_i\ln C + h_i + p_i( \partial_j h_j + h_j \partial_j\ln C)=0$, but this cannot be achieved because we may adjust $C$ to compensate terms along the $h_i$ or the $p_i$ direction, but not both simultaneously. This suggests ambiguities in the ordering of auxiliary operators plague the algebraic approach to models with anisotropies coupling directly to $\hqi$ in $\hxi$ --- a situation analogous to that reported at the end of Sec.\ \ref{sec:symm-x-non-pert}.

\subsubsection{Perturbative approach}

As in Sec.\ \ref{sec:symm-x-pert}, we start with a single anisotropy symbolically denoted by $a$, corresponding to \textit{one} anisotropic term out of the three possibilities in (\ref{single-anisotropies}), so that $\fij=\dij+f_{ij}(\bp)$ with $f_{ij} \sim \oo{a}$. Changing the ordering of auxiliary operators in $\hxi$ from $\fij\hqj$ to $\hqj \fij$ adds the term $[\fij,\hqj]=-\ih\partial_j f_{ij}$ which is $\oo{a}$ as well. Meanwhile, a canonical transformation with $C(\bp)$ adds $-\ih \fij\partial_j\ln C$, which reduces to $-\ih \partial_i C+\oo{a^2}$ for $C=1+\oo{a}$. Thus, any $C$ satisfying $\partial_j f_{ij} + \partial_i C = \oo{a^2}$ renders such reordering unobservable to $\oo{a}$. For the $\fij$ just described, it is always possible to find such $C$, \eg $C\propto \dij \fij$. In particular, this is also true for models based on $h_j$ with a single anisotropic term. Intermediate reordering is found to be unobservable on the same grounds.

Approximate models with multiple anisotropies may be handled in similar fashion, but extra care is needed since the anisotropies usually have differing dimensionality. Restricting our discussion to the anisotropies on (\ref{particular-anisotropies}) for clarity, we notice one extra power on vector anisotropies should be retained in comparison to tensor ones whenever considering approximations to some specific order. To illustrate what this means for reordering, next we provide an example.

Consider $\fij = (1+\pbp)\dij + p_i c_j$. Since $[c_i]=[p]^{-1}$ and $[\beta_{ij}]=[p]^{-2}$, our analysis will be to first order on $\beta_{ij}$ but to second order on $c_i$. The canonically transformed position operator is constructed from (\ref{hxi-prime}) and we propose
\be
\label{perturbativeC}
C(\bp)=1 +\kappa_1 \pbp +\kappa_2 \pc +\kappa_3 (\pc)^2 +\kappa_4 c^2 \bp^2,
\ee
with adjustable constants $\kappa_i$ to render the reordering $\fij\hqj\to\hqj \fij$ unobservable. This is achieved to the desired order if $C(\bp)$ satisfies (\ref{condition-reordering}), \ie such that
\be
\label{perturbativaexample}
\partial_j \fij = 2(\pb)_i + c_i
\ee
and
\begin{align}
\label{perturbativecounter}
& \fij\partial_j\ln C
\nonumber\\
&
= C^{-1}\{ [(1+\pbp)\dij + p_i c_j]
\nonumber\\
& \quad 
\qquad \times [2\kappa_1 (\pb)_j + (\kappa_2 +2\kappa_3 \pc) c_j + 2\kappa_4 c^2\hpi] \}
\nonumber\\
&
\approx 2\kappa_1 (\pb)_i + [\kappa_2 \!-\! (\kappa_2 \!-\! 2\kappa_3) \pc] c_i + (\kappa_2 \!+\! 2\kappa_4) c^2 p_i
\end{align}
have zero-sum to $\oo{\bb,c^2}$. In passing, notice $C^{-1}$ above was not completely disregarded because it contributes a term of $\oo{c^2}$. What is new here is the momentum dependent factor multiplying $c_i$ and the term proportional to $\hpi$ in the last equation above. It traces back to the need of keeping terms of $\oo{c^2}$ as we keep others of $\oo{\bb}$; this is common for perturbative models with anisotropies of different tensorial nature. These momentum-dependent factors need to vanish in face of (\ref{perturbativaexample}) and we remark our ability to do so comes from introducing suitable $\oo{c^2}$ terms in (\ref{perturbativeC}). Thus, setting $\kappa_1=-1$ and $\kappa_2=2\kappa_3=-2\kappa_4=-1$ we confirm the reordering $\fij\hqj\to\hqj \fij$ is unobservable to this order. Analogously, any intermediate reordering can be compensated by a canonical transformation as well.

\section{Models with overlapping features}
\label{sec:overlap}

In this section we summarize the models featuring invariance of commutators (\ref{deformed-commxp}) and (\ref{deformed-commxx}) under canonical transformations and exhibiting no issues regarding the ordering of auxiliary operators in the definition of the position operator. Brief discussion of interesting features of each model is also provided. Here we focus only on GUP models in anisotropic space because those in isotropic space have all these features for any reasonable $C(p)$. Also, perturbative results are not discussed here since we found no major obstacle on its realization for the models considered in this paper. As a reminder, we keep the choice of representing the momentum operator as the usual multiplicative operator; hence, all canonical transformations we mention next are based on $C=C(\bp)$ and, when not given explicitly, are assumed to behave like $C\to1$ as $\hp\to0$.

\subsection{\texorpdfstring{$\fij = f(\bp)\dij$}{} }

The symmetric position operator in this case is given by
\be
\hxi = f(\bp)\hqi + \tfrac{1}{2}\ih\partial_i f(\bp)
\ee
and its canonically transformed version is
\be
\hxi' = \hxi - \ih f(\bp)\partial_i \ln\C.
\ee
The ordering $f\hqi$ or $\hqi f$ is immaterial as the second is equivalent to setting $\hxi'$ with the choice $C\propto f^{-1}$. Moving $\hqi$ inside $f(\bp)$ in between momentum factors is a different matter: only for $f=1+\w{f}$ as in (\ref{particular-anisotropies}) there is $C\propto f^{-n}$ rendering the reordering unobservable for suitable $n$. Only for $f$ with this specific structure we securely say reordering of auxiliary operators is totally unobservable. Apart from this, in what follows we consider $f$ as general as possible.

The commutator of the position and momentum operators is invariant under general $C(\bp)$ and reads
\be
\commij = \ih f\dij,
\ee
but the commutator of position operators,
\be
\commxx = -\ih\varepsilon_{ijk} (\hxi \partial_j f - \hxj \partial_i f),
\ee
on the other hand, is invariant only for $C$ proportional to some function of $f$. 

Notice the last commutator above reduces to $-\ih\varepsilon_{ijk}(f f'/\hp) L_k$ for isotropic $f(p)$ if one identifies $\mbf{L} = \bq\times\bp =  f^{-1} \bx\times\bp$ as the operator satisfying the conventional angular momentum algebra; namely, $[\hpi,L_j]=\ih\varepsilon_{ijk}\hp_k$, $[\hxi,L_j]=\ih\varepsilon_{ijk}\hx_k$, and $[L_i,L_j]=\ih\varepsilon_{ijk}L_k$. Proposing a similar definition for the anisotropic case, one finds instead $\mbf{L} = \bq\times\bp =  f^{-1} \bx\times\bp -\tfrac{1}{2}\ih f^{-1} \nabla f \times \bp$, illustrating the coupling of the momentum $\bp$ to a preferred direction $\nabla f$ in space indeed. Notice $\mbf{L}$ still commutes with the free particle Hamiltonian $H=p^2/2m$, indicating the angular momentum is conserved despite the spatial anisotropy.

\subsection{\texorpdfstring{$\fij = \dij + g_i(\bp)\hpj$}{} }

Our conclusions are less general for models with $\fij = \dij +g_i(\bp)\hpj$, for which we are led to consider only a single anisotropic term at a time, the setting $g_i=\w{g}_i$, and corresponding anisotropy $g_\ione$ to be symmetric under exchange of any pair of indices --- for instance, $g_i$ may stand for $\alpha'_i$ or $(\pb')_i$ with symmetric $\beta'_{ij}$ in (\ref{particular-anisotropies}).

The symmetric position operator and its canonically transformed version are, respectively,
\be
\hxi = \hqi + \w{g}_i \left( \pdotq + \frac{d+2}{2}\ih \right),
\ee
\be
\hxi' = \hxi -\ih [ \partial_i\ln\C + \w{g}_i \bp\cdot\nabla \ln\C ].
\ee
Setting $C=(1+\w{g})^{n}$ gives $\hxi'$ equivalent to defining $\hxi$ with auxiliary operators ordered in different fashion. For the basic commutators, one finds
\be
\commij = \ih[\dij + \w{g}_i\hpj ],
\ee
\be
\commxx = \ih\varepsilon_{ijk}( \hxi \w{g}_j - \hxj \w{g}_i ).
\ee
The last commutator is invariant only for canonical transformations with $C$ proportional to any reasonable function of $\w{g}$ added by a constant factor.

Defining the operator $L_k = \ijk \hqi\hpj$ satisfying the conventional angular momentum algebra, we find
\be\label{g-angular-momentum}
L_k = \ijk \hxi\hpj - \ijk \w{g}_i \hpj \frac{1}{1+\w{g}} \left( \bp \cdot \bx + \frac{d+2}{2}\ih \right),
\ee
where the anisotropic factor $\w{g}_i$ couples to the momentum by means of a cross product, similarly to the previous case just discussed. On the other hand, the free particle Hamiltonian does not commute with $L_k$,
\be
[H,L_k] \propto \ijk \w{g}_i \hpj \frac{p^2}{1+\w{g}},
\ee
revealing the angular momentum is no longer conserved, except for particle propagation parallel to $\w{g}_i$.

\subsection{\texorpdfstring{$\fij = f(\bp)\dij + g_i(\bp)\hpj$}{} }

We have found the case of $f$ and $g_i$ both with a single anisotropic term can be considered simultaneously when the anisotropies are of same tensor rank and taken as proportional to each other. One example is $f=1+\pa$ and $g_i \propto \alpha_i$; another is $f=1+\pbp$ and $g_i \propto (\pb)_i$. For greater generality, next we consider $f$ and $g_i$ as in (\ref{single-anisotropies}).

For $f=1+\w{f}$ we set $g_i=\w{g}_i$ with same $d$ so both have anisotropies of the same tensor rank. Also, the two anisotropies are taken as satisfying $g_\ionetwo = \kappa f_\ionetwo$. The associated position operators are
\be
\hxi = (1+\w{f})\hqi + \w{f}_i [ \kappa\pdotq + (\kappa + \tfrac{1}{2}\kappa d + \tfrac{1}{2}d)\ih ],
\ee
\be
\hxi'=\hxi -\ih [ (1+\w{f})\partial_i\ln\C + \kappa \w{f}_i \bp\cdot\nabla \ln\C ].
\ee
Here the transformation $C=[1+(1+\kappa)\w{f}]^{\frac{n}{1+\kappa}}$ with adjustable $n$ sets a model equivalent to others with different ordering of auxiliary operators. At last, the commutators
\be\label{summary-f-g}
\commij = \ih [(1+\w{f})\dij + \kappa\w{f}_i \hpj],
\ee
\be
\commxx = \ih\varepsilon_{ijk}\varepsilon_{kmn}\frac{ d-\kappa+(d-\kappa+\kappa d)\w{f}}{1+\w{f}} \w{f}_m \hx_n.
\ee
are simultaneously invariant under canonical transformations for any suitable $C$ that is a function of $\w{f}$. The operator identified as the angular momentum is found to be structurally analogous to (\ref{g-angular-momentum}), suggesting it is generally not conserved even for the free particle propagation also in this case.

An interesting feature of the particular case $\kappa=d$ is the commutativity of position operators to $\oo{\widehat{f}}$. To this order, translation by any amount $\mbf{a}$ is a symmetry of the theory because $\commxx=\oo{\widehat{f}^2}$ and $\commij$ are invariant under the unitary operator $U(\bp)$ implementing $U \bx U^\dagger = \bx + \mbf{a}$. This operator is found to be $U = \exp(-\ih\,\mbf{a}\cdot\mbf{T})$, where $\mbf{T}=\bp/(1+\w{f})$ is the generator of translations satisfying $[\hxi,T_j]=\ih\dij+\oo{\widehat{f}^2}$. Acting on eigenstates of $T_i$ with corresponding eigenvalues $\rho_i$, this last commutator reads $[\hxi,\rho_j]=\ih\dij+\oo{\widehat{f}^2}$, revealing the alternative representation $\hxi=\ih\partial_{\rho_i}$ and $\hpi=(1+\w{f})\rho_i$ is now found to satisfy the commutator (\ref{summary-f-g}) to $\oo{\widehat{f}}$.

As an illustration, consider the model with $d=2$, for which the commutator of position and momentum is
\be
\commij = (1+\pbp)\dij + (\pb')_i \hpj.
\ee
This model provides a generalization of the popular one considered by Kempf in \cite{kempf97}, recovered in the isotropic space limit, $\beta_{ik}\to\beta\delta_{ik}$ and $\beta'_{ik}\to\beta'\delta_{ik}$. Setting $\bb'=\kappa\bb$, the relevant commutators are invariant under suitable canonical transformations and there are no issues regarding the ordering of auxiliary operators, as already discussed. The commutator of position operators simplifies to
\be
\commxx = \ih\varepsilon_{ijk}\varepsilon_{kmn}\frac{ 2-\kappa+(2+\kappa)\pbp}{1+\pbp} (\pb)_m \hx_n
\ee
and actually vanishes to $\oo{\bb}$ for $\bb'=2\bb$. This particular case was considered in \cite{nigup} and it was found that searches for annual variations in high precision measurements of the 1S-2S hydrogen transition frequency bounds the trace of $\bb$ at the $10^{-8}\,\text{GeV}^{-2}$ level in the laboratory frame. Still for the case $\bb'=2\bb$, the commutator of position and momentum reads
\be\label{anisotropic-kempf}
\commij=\ih[(1+\pbp)\dij+2(\pb)_i\hpj]
\ee
and is realized, to any order in $\bb$, for a representation based on eigenstates of the momentum,
\begin{align}
\hxi & = \ih(1+\pbp)\partial_{p_i} + \ih(\pb)_i (2\bp\cdot\partial_{\bp_i} + 5),
\nonumber\\
\hpi & =\text{ usual multiplicative operator},
\end{align}
where we used (\ref{hxi}) to compute $x_i$. Alternatively, the above commutator is realized to $\oo{\bb}$ for the representation based on eigenstates of the generator of translations,
\begin{align}
\hxi & = \ih\partial_{\rho_i},
\nonumber\\
\hpi & = \rho_i (1+\mbf{\rho}\cdot\bb\cdot\mbf{\rho}).
\end{align}
Although the equivalence of the two representations is valid only to $\oo{\bb}$, it suggests there may exist models featuring exact equivalence even in the anisotropic scenario; this is true indeed, as we see next.

\subsection{Commutative models}
\label{sec:commutative}

For $f$ and $g_i$ containing each a \textit{single} anisotropic term, as in the previous section, we see there are particular models with \textit{approximately} commutative position operators. Incidentally, hint at the existence of a further general result is provided noting the isotropic space case, $F_{ij} = f(p)\dij + g(p)p_i p_j$, features
\be
\commxx=0 \qquad \Longleftrightarrow \qquad gp=\frac{f\partial_p f}{f-p\partial_p f}
\ee
valid to all orders \cite{kempf97,igup}. The anisotropic version of this \textit{commutativity condition} is found for the case $F_{ij} = f(\bp)\dij+g_i(\bp)p_j$ and given by
\be\label{gup-comm}
\commxx = 0 \qquad \Longleftrightarrow \qquad g_i = \frac{f \partial_{\hpi} f}{f - \bp\cdot\partial_{\bp} f},
\ee
also valid to all orders \cite{nigup}.\footnote{For the slightly more general case $F_{ij} = f(\bp)\dij + g_i(\bp) h_j(\bp)$ we have found there is no condition allowing for exact commutativity of position coordinates unless $h_j(\bp)=p_j$. For the general case $F_{ij}$ we were not able to arrive at any concrete conclusion.} It enforces, in particular, anisotropies $g_i$ and $\partial_{\hpi} f$ to be parallel.

For such commutative GUP model in anisotropic space, setting $g_i$ as given by (\ref{gup-comm}) and using (\ref{hxi}) for $x_i$, we find the commutator of position and momentum is realized in momentum space representation for
\begin{align}
\hxi & = \ih f \partial_{p_i} + \ih g_i( \bp\cdot\partial_{\bp} + \tfrac{3}{2}) + \tfrac{1}{2}\ih (\partial_{\bp_i}f + \bp\cdot\partial_{\bp} g_i),
\nonumber\\
\hpi & =\text{ usual multiplicative operator}.
\end{align}
The alternative representation on the spectral basis $|\mbf{\rho}\ra$ of the generator of translations $\mbf{T}$ is
\begin{align}
\hxi & = \ih\partial_{\rho_i},
\nonumber\\
\hpi & = \rho_i f(\bp(\mbf{\rho})),
\end{align}
where the last is an implicit equation for $\hpi$ as a function of the eigenvalues $\rho_i$ of $T_i$. This time, the equivalence of the two representations holds \textit{exactly} and, for this reason, we expect any reordering of $\hqi$ and $\hpi$ in $\hxi(\bq,\bp)$ to be unobservable since the second representation above automatically features no issue regarding ordering of auxiliary operators.

\section{Concluding remarks}
\label{sec:conclusion}

Motivated by recent work relating GUP models in anisotropic space and Lorentz symmetry violation \cite{nigup}, in this paper we identified GUP models satisfying: (i) invariance of $\commij$ and $\commxx$ under canonical transformations, and (ii) $x_i$, when expressed as functions of the auxiliary operator $q_i$ satisfying $[q_i,p_j]=\ih\dij$, do not physically depend on the particular ordering of $q_i$ and $p_i$ in its definition. This identification is specially important as a first step in sorting out which of the diverse possibilities for GUP in anisotropic space allow for a consistent description under the algebraic approach to GUP.

For the anisotropic models considered here, a seemly general conclusion is that neither of the mentioned criteria is satisfied whenever space anisotropies (\eg background vector $c_i$) couple to the auxiliary operator $q_i$ in the expression for $x_i$ (\eg $x_i \sim q_i + p_i \bc\cdot\bq$). This conclusion suggests that the GUP model $\commij = \ih[f(\bp)\dij+p_i h_j(\bp)]$ is not straightforwardly described within the algebraic approach by setting $x_i \sim f q_i + p_i \mbf{h}\cdot\bx$.

In contrast, many models based on $\commij = \ih[f(\bp)\dij+g_i(\bp) p_j]$ are found to satisfy both criteria and, for these, anisotropies never appear as dot products with $q_i$ since $x_i \sim f q_i + g_i \pdotq$. The algebraic approach to GUP seems suitable for describing these models. A limiting aspect of this conclusion is that it seemly applies only to particular $f$ and $g_i$, usually assuming existence of no more than a single kind of anisotropy. A notable exception is the commutative model derived for $f$ and $g_i$ related by the condition (\ref{gup-comm}). This last finding underpins the investigation performed on \cite{nigup} focused on the physics of such commutative model and should go side by side with it.

Considering future works, physics from models satisfying the two criteria is largely unexplored and ought to be further investigated. For instance, the approximately commutative model based on (\ref{anisotropic-kempf}) is found on \cite{nigup} to predict annual variations of the 1S-2S hydrogen transition frequency. Predictions from other models in Sec.\ \ref{sec:overlap} might be learned after comparison to the Lorentz-violating Standard Model Extension, in a similar way to what was done in \cite{nigup,gup-sme}. It might also be interesting to investigate whether these models feature minimal length uncertainty and, if so, how the space anisotropy affects it --- \eg  the commutator (\ref{anisotropic-kempf}) implies a direction-dependent uncertainty of $(\Delta x_i)_\text{min} \sim \sqrt{\beta_i}$ in the particular case $\boldsymbol{\beta} = \text{diag}(\beta_1,\beta_2,\beta_3)$, but reaching a definite conclusion in more general situations may be challenging. Other potentially interesting routes for investigation of anisotropic effects include the connection of GUP and entropy \cite{tsallis,entropy}, GUP from curved momentum space \cite{wagner-anisotropic}, and the path integral formulation \cite{path-integral}.

At last, our results suggest two questions related to canonical transformations that might be worth pursuing. The first deals with the Stone-von Neumann theorem. For conventional quantum mechanics, this theorem ensures that representations of the canonical commutation relation are all unitarily equivalent to each other. For the quantum mechanics considered in this work, the validity of the theorem may not be straightforward since $\commij \neq \ih\dij$ and $\commxx\neq0$. If the theorem is not valid in some circumstances, would this result be related to the GUP model $\commij = \ih[f(\bp)\dij+p_i h_j(\bp)]$ found to be in tension with the algebraic approach to GUP? The second question is related to the extension of GUP to quantum field theory (QFT). Since the existence of unitarily inequivalent representations of canonical commutation relations is common in QFT and often physically meaningful \cite{jizba-book}, it should be interesting to learn whether the canonical transformations considered in Sec.\ \ref{sec:overlap} have a non-unitary QFT counterpart and, if so, what is the physics behind it.

\section{Acknowledgments}

The author wish to thank Pasquale Bosso for many pertinent remarks and also the two anonymous referees for the very relevant feedback.

%
\bibliography{references}

\end{document}